\documentclass{pasj00}

\begin{document}
\SetRunningHead{S.\ Yamauchi et al.}{Evidence for Recombining Plasma in the Supernova Remnant G346.6$-$0.2}
\Received{2012/05/29}
\Accepted{2012/8/23}

\title{Evidence for Recombining Plasma in the Supernova Remnant G346.6$-$0.2}

 \author{%
   Shigeo \textsc{Yamauchi}\altaffilmark{1}, Masayoshi \textsc{Nobukawa}\altaffilmark{2},
   Katsuji \textsc{Koyama}\altaffilmark{2,3},   
   and 
Manami \textsc{Yonemori}\altaffilmark{3}
}
 \altaffiltext{1}{Department of Physics, Faculty of Science, Nara Women's University, 
Kitauoyanishi-machi, Nara 630-8506}
\email{yamauchi@cc.nara-wu.ac.jp}
\altaffiltext{2}{Department of Physics, Graduate School of Science, Kyoto University, \\
Kitashirakawa-oiwake-cho, Sakyo-ku, Kyoto 606-8502}
\altaffiltext{3}{
Department of Earth and Space Science, Graduate School of Science, Osaka University, \\
1-1 Machikaneyama-cho, Toyonaka, Osaka 560-0043}


%

\KeyWords{ISM: individual (G346.6$-$0.2) --- ISM: supernova remnants --- X-rays: ISM --- X-rays: spectra} 

\maketitle

\begin{abstract}
We present the Suzaku results of the supernova remnant (SNR) G346.6$-$0.2.
The X-ray emission has a center-filled morphology with the size of 6$'$$\times$8$'$ within the radio shell. 
Neither an ionization equilibrium nor non-equilibrium (ionizing) plasma can reproduce the spectra 
remaining shoulder-like residuals in the 2--4 keV band.
These structures are possibly due to recombination of free electrons to the K-shell of 
He-like Si and S. The X-ray 
spectra are well fitted with a plasma model in a recombination dominant phase. 
We propose that the 
plasma was in nearly full ionized state at high temperature of $\sim$5 keV, then 
the plasma changed to a recombining phase due to selective cooling of 
electrons to lower temperature of $\sim$0.3 keV. 
G346.6$-$0.2 would be in an epoch of the recombining phase.
\end{abstract}

\section{Introduction}

G346.6$-$0.2 is a supernova remnant (SNR) discovered in the radio band 
\citep{Clark1975}. The radio image shows a shell structure with the size of
$\sim$8$'$ \citep{Clark1975,Dubner1993, Whiteoak1996}. Flux densities 
at 408 MHz, 843 MHz, 1.47 GHz, and 5 GHz were measured to be 14.9 Jy, 8.7 Jy, 8.1 Jy, 
and 4.3 Jy, respectively (\cite{Green2009} and references therein), 
then the spectral index was estimated to be 0.5 \citep{Clark1975,Green2009}. 
The SNR would be interacted with a molecular cloud because an OH maser was 
found \citep{Green1997}.

The Galactic plane survey project with ASCA discovered a faint X-ray emission from 
G346.6$-$0.2 for the first time \citep{Yamauchi2008}. 
The ASCA GIS image showed the diffuse X-ray morphology in the radio shell. 
The X-ray spectrum was represented 
by either a thermal plasma model with a temperature of $\sim$1.6 keV or a power-law model 
with a photon index of $\sim$3.7. 
The absorption was as large as $N_{\rm H}=$(2--3)$\times10^{22}$ cm$^{-2}$ 
\citep{Yamauchi2008}, which suggested that the SNR is located at a long distance, 
possibly in the Galactic inner disk or further. 
No further quantitative constraint was available with the ASCA data 
due to the limited photon statistics. 
G346.6$-$0.2 was then observed with Suzaku. 
\citet{Sezer2011} found strong emission lines of Si and S, 
and fitted the X-ray spectrum with a model of a power-law 
(photon index $\sim$0.6) plus a non-equilibrium ionization (NEI) (ionizing) plasma. 
They predicted that the strong Si and S lines are due to an ejecta-dominated plasma 
which originated from a Type Ia supernova explosion, 
and the power-law component is regarded as synchrotron emission.  

Recently,  
strong radiative recombination continua (RRCs) have been discovered 
in the X-ray spectra of five mixed-morphology (MM) SNRs 
\citep{Yamaguchi2009,Ozawa2009,Ohnishi2011, Sawada2012, Uchida2012}.
The RRC originates from radiative transitions of free electrons to the K-shell of ions, 
a sign of a recombination dominant plasma (RP).
In the residuals of the NEI model fit for G346.6$-$0.2 in \citet{Sezer2011}, 
we see a similar structure to the RRC.

G346.6$-$0.2 is located on the Galactic ridge, where a strong X-ray emission, called the 
Galactic Ridge X-ray Emission (GRXE), is prevailing. 
However, the previous data reduction and analyses did not properly take account of 
the GRXE as a major background for the faint and diffuse source. 
We, therefore, revisited the Suzaku data and 
performed the data reduction, spectral construction, and spectral analysis 
paying particular concerns on the subtraction of the GRXE. 
We then discovered evidence for the recombining plasma
from this MM SNR for the first time.

\section{Observation and Data Reduction}
\begin{figure*}
  \begin{center}
    \FigureFile(17cm,8cm){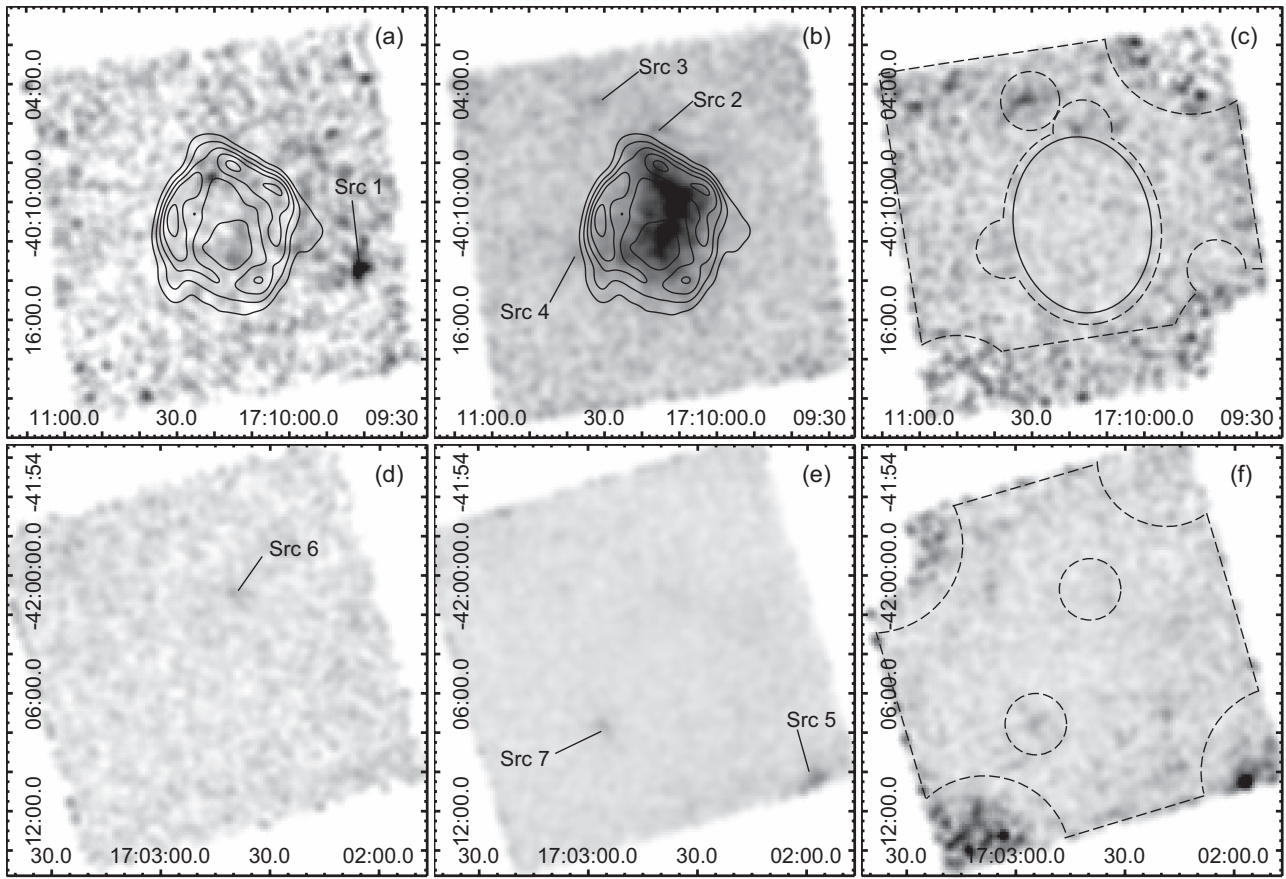}
  \end{center}
  \caption{
Upper panel: 
XIS images of G346.6$-$0.2 in the 0.5--1 (a), 1--5 (b), and 5--8 keV (c) energy bands (gray scale). 
The coordinates are J2000.0. 
The radio map at 843 MHz using the Molonglo Observatory 
Synthesis Telescope (MOST) is displayed by the solid contours in (a) and (b) \citep{Whiteoak1996}.
The X-ray images from XIS\,0, 1, and 3 were co-added. After the subtraction of the  
Non X-ray background (NXB), the vignetting  corrections were made. 
The images were smoothed 
with a Gaussian distribution with the kernel of $\sigma$=24$''$.
The intensity levels of the X-ray and radio bands are linearly spaced.
Point-like sources are labeled as Src~1--Src~4.
The background region (BGD-a) is shown by the dashed lines excluding the point-like sources 
(the dashed circles) in (c).   
The solid ellipse in (c) shows the source region. 
Lower panel: Same as (a)-(c), but the X-ray image of the background sky on a nearby 
source-free Galactic ridge.  
Src 5--7 are point-like sources.  
The background region (BGD-b) is shown by the dashed lines excluding the point-like sources 
(the dashed circles) in (f). 
}\label{fig:sample}
\end{figure*}

Suzaku \citep{Mitsuda2007} carried out the Galactic center/plane mapping project 
with the CCD cameras (XIS, \cite{Koyama2007}) 
placed at the focal planes of the thin foil X-ray Telescopes (XRT, \cite{Serlemitsos2007}). 
The SNR G346.6$-$0.2 was observed on 2009 October 7--9 (Obs. ID 504096010).
The pointing position was 
($l$, $b$)=(346$^{\circ}$.63, $-$0$^{\circ}$.22). 

G346.6$-$0.2 is a faint X-ray SNR, located toward the inner Galactic disk. 
Therefore, the contribution of the GRXE, 
particular in the hard X-ray band above 5~keV, has a large impact on the source spectrum.  
Although the surface brightness of the GRXE is nearly constant along the 
Galactic plane within the range of a few degree of the Galactic longitude, the latitudinal 
variation is large, which is given by an exponential function with the scale height of 
$\sim\timeform{0.D5}$ (e.g., \cite{Koyama1986,Yamauchi1993,Kaneda1997b}). 
We, therefore, selected two background regions for the spectrum of G346.6$-$0.2; 
one is the surrounding region of G346.6$-$0.2 in the same field of view (FOV) (hereafter BGD-a) 
and the other is a nearby region at 
($l$, $b$)=(344$^{\circ}$.26, $-$0$^{\circ}$.22) (Obs. ID 502049010; hereafter BGD-b), 
at the same Galactic latitude but $\timeform{1D}$ away from G346.6$-$0.2 in longitude.

XIS sensor-1 (XIS\,1) is a back-side illuminated (BI) CCD, while the other three XIS sensors 
(XIS\,0, 2, and 3) are front-side illuminated (FI) CCDs.
The FOV of the XIS is 17.8$'$$\times$17.8$'$.
Since one of the FIs (XIS\,2) turned dysfunctional in 2006 November\footnote{http://www.astro.isas.ac.jp/suzaku/news/2006/1123/}, 
we used the data obtained with the other CCD cameras (XIS\,0, 1, and 3). 
A small fraction of the XIS0 area was not used because of the data damage 
due possibly to an impact of micro-meteorite on 2009 June 23\footnote{http://www.astro.isas.ac.jp/suzaku/news/2009/0702/}.
The XIS was operated in the normal clocking mode.
The degraded spectral resolution due to the radiation of cosmic particles 4 years after the 
launch was restored by the spaced-row charge injection (SCI) technique.
Details of the SCI technique are given in \citet{Nakajima2008} and \citet{Uchiyama2009}.

Data reduction and analysis were made with the HEAsoft version 6.11, 
SPEX \citep{Kaastra1996} version 2.02.04, and the processed data version 2.4. 
The XIS pulse-height data for each X-ray event were converted to 
Pulse Invariant (PI) channels using the {\tt xispi} software 
and the calibration database version 2011-11-09.
We rejected the data taken at the South Atlantic Anomaly, 
during the earth occultation, and at the low elevation angle 
from the earth rim of $<5^{\circ}$ (night earth) and $<20^{\circ}$ (day earth).  
The exposure times after these screenings were 56.8 and 215.7 ks 
for G346.6$-$0.2/BGD-a and BGD-b fields, respectively. 

In the following data analysis, we subtracted the non-X-ray background (NXB),
which was constructed from the night earth data generated by {\tt xisnxbgen} \citep{Tawa2008} 
in the HEAsoft package.

\section{Analysis and Results}

\subsection{X-Ray Image}

Figure 1 shows the vignetting corrected X-ray images for the 
G346.6$-$0.2/BGD-a (1a-1c) and BGD-b (1d-1f) fields, in the 0.5--1, 1--5, and 5--8 keV energy bands. 
The solid contours in figures 1a and 1b are the radio band image at the 843 MHz \citep{Whiteoak1996}.  
In order to increase X-ray photon statistics, we co-added all the data of XIS\,0, 1, and 3.  
The images contain all the X-rays including the GRXE and the cosmic X-ray background (CXB).  
Above the X-ray fluxes of these backgrounds,  
diffuse X-rays from G346.6$-$0.2 are found only in the 1--5 keV 
band image (figure 1b); 
no significant X-ray is found in the 0.5--1 and 5--8 keV bands (figures 1a and 1c). 
We see a center-filled X-ray emission within the shell of the radio emission. 
In addition, four and three faint point-like sources are found in the G346.6$-$0.2 (Src 1--4) and 
BGD-b field (Src 5--7), respectively.

\subsection{X-Ray Spectra}
\begin{figure}
  \begin{center}
   \FigureFile(8cm,8cm){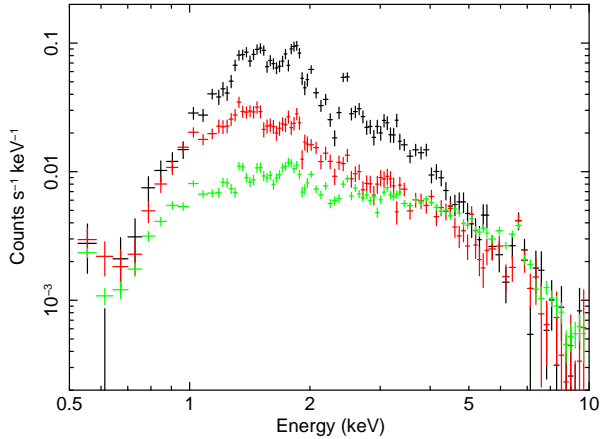}
  \end{center}
  \caption{
Comparison of X-ray spectra of the source (black), the backgrounds 
of BGD-a (red), and BGD-b (green), after subtracting the NXB.
 Only the XIS\,0 data are plotted for brevity.
 }\label{fig:sample}
\end{figure}

The XIS spectra of G346.6$-$0.2, BGD-a, and BGD-b were extracted from the solid 
ellipse in figure 1c, the dashed lines in figures 1c, and 1f, respectively.  
For these data, we excluded regions of the point-like sources (Src 1--7) and 
the calibration sources at the field corners.
The source and background spectra are shown in figure 2.
The hard X-ray flux above $\sim$5 keV is essentially comparable with each other, 
as is also noted in the hard X-ray band image of figure~1.
The hard X-rays are likely due to the sum of the CXB and the
GRXE. The Fe K-line emission at 6.7 keV is due to the GRXE, and the line intensities 
are also comparable between the source and two background spectra. 
We, therefore, conclude that after the subtraction of the proper background of 
the GRXE and the CXB, G346.6$-$0.2 should have no significant X-rays above $\sim$5~keV.

The BGD-a spectrum shows an X-ray excess over BGD-b in the low energy band below $\sim$3~keV. 
This implies that a local diffuse emission is prevailing around G346.6$-$0.2. 
We discuss this local diffuse emission in section 4.2.
BGD-a should be a better background for G346.6$-$0.2 because of the close vicinity, 
but photon statistics are limited due to less exposure time of 56.8 ks and off-axis vignetting effect. 
BGD-b, on the other hand, provides better photon statistics due to longer exposure of 215.7 ks 
and larger correcting area than BGD-a. 
We, therefore, made two source spectra by the subtraction of each background, BGD-a or BGD-b. 
The exposure times and vignetting effects of these backgrounds were 
corrected by the method shown by \citet{Hyodo2008}.

For the spectral analysis, the Response files, Redistribution Matrix Files (RMFs) and 
Ancillary Response Files (ARFs), were made using {\tt xisrmfgen} and {\tt xissimarfgen}, respectively, 
in the HEAsoft package. 
The ARFs were created assuming the observed image to the photon distribution.

\subsubsection{G346.6$-$0.2}

\begin{figure*}
  \begin{center}
   \FigureFile(16cm,16cm){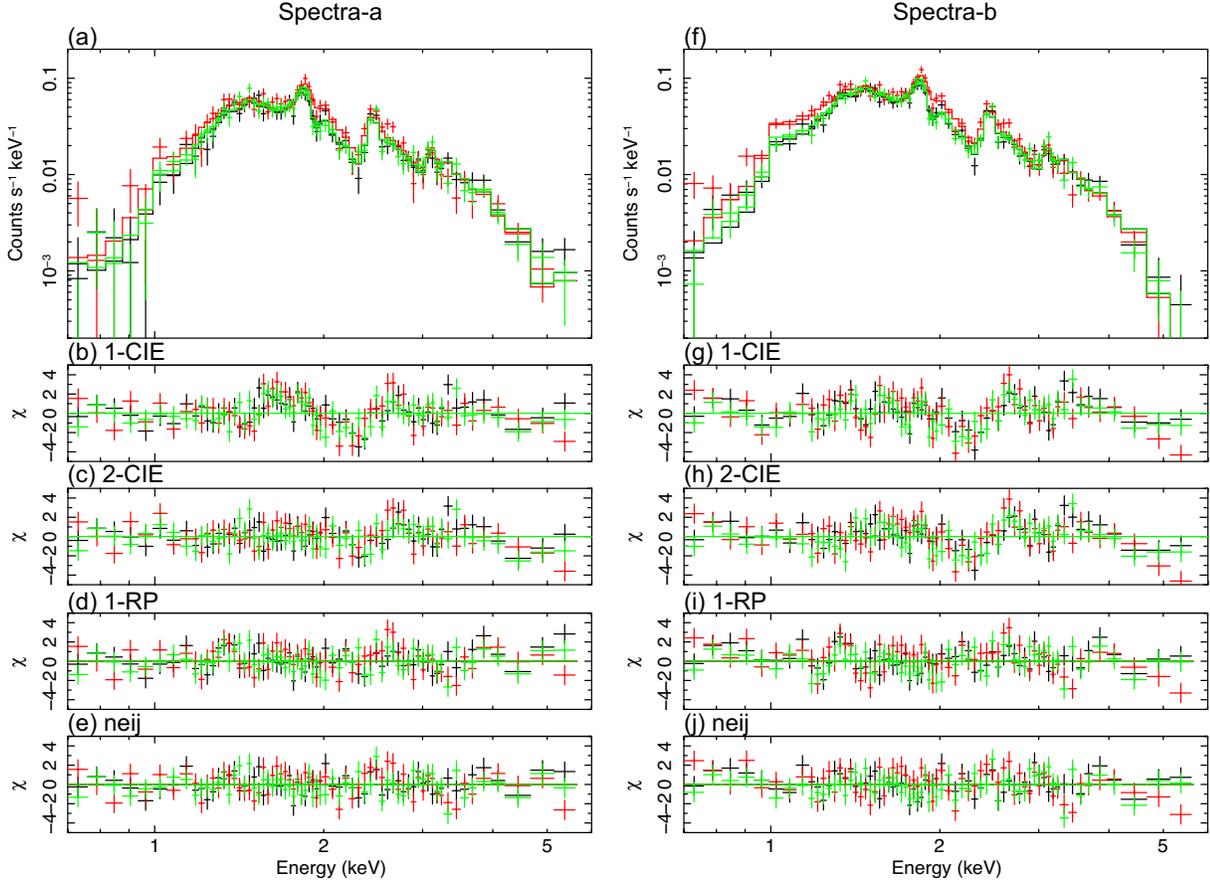}
  \end{center}
  \caption{
X-ray spectra of G346.6$-$0.2  obtained with  the Suzaku XIS and the residuals from the best-fit 
model (see text). The black, red, and green colors show XIS\,0, 1, and 3 data, respectively.
The left panels are the results of  spectra-a (BGD-a subtracted spectra),  
while the right panels are those of spectra-b (BGD-b subtracted spectra).
The histograms in (a) and (f) are the best-fit {\tt neij} models (see table 1), while 
the residuals from the models of 1-CIE, 2-CIE, 1-RP and {\tt neij} are shown
in the histograms of (b), (c), (d) and (e) (for spectra-a), and (g), (h), (i) and (j) (for spectra-b), respectively.
}\label{fig:sample}
\end{figure*}

%
\begin{table*}[t]
\caption{The best-fit parameters derived from a spectral analysis.$^{\ast}$}
\begin{center}
\begin{tabular}{lcccc} \hline  \\ [-6pt]
Parameter & \multicolumn{4}{c}{Value} \\
Model  &  XSPEC: vapec   & XSPEC: 2 vapec & SPEX: cie & SPEX: neij\\
&  1-CIE    & 2-CIE & 1-RP & \\
\hline \\ [-6pt]
\multicolumn{5}{c}{Spectra-a}\\
\hline \\ [-6pt]
$N_{\rm H}$ ($\times10^{22}$ cm$^{-2}$)  &
2.9$^{+0.3}_{-0.4}$  & 3.8$^{+0.3}_{-0.5}$ & 1.9$^{+0.1}_{-0.1}$ & 2.3$^{+0.1}_{-0.1}$\\
$kT_{\rm e}$ (keV)  & 
1.01$^{+0.04}_{-0.04}$  & 0.18$^{+0.04}_{-0.02}$ & 0.37$^{+0.04}_{-0.03}$ & --- \\
   &     ---            & 1.19$^{+0.15}_{-0.10}$ &   ---  & --- \\
$kT_{\rm z}$/$kT_{\rm e}$ & ---  & --- & 2.9$^{+0.4}_{-0.3}$ & ---\\
$kT_{\rm e1}$ (keV)  &  --- & --- & --- & 5 (fixed)\\
$kT_{\rm e2}$ (keV)  &  --- & --- & --- & 0.30$^{+0.03}_{-0.01}$\\
$n_{\rm e}t^{\dag}$ ($\times10^{11}$ cm$^{-3}$ s) & 
---  & --- & --- & 4.8$^{+0.1}_{-0.4}$\\
Mg$^{\dag}$   & 
1.7$^{+0.9}_{-0.6}$  & 0.4$^{+0.4}_{-0.3}$ & 0.2$^{+0.2}_{-0.2}$ & 0.4$^{+0.1}_{-0.2}$ \\
Si$^{\ddag}$   & 
1.1$^{+0.8}_{-0.5}$  & 1.7$^{+1.0}_{-0.5}$ & 0.8$^{+0.1}_{-0.2}$  & 0.6$^{+0.2}_{-0.1}$\\
S$^{\ddag}$     & 
1.3$^{+0.5}_{-0.4}$  & 2.7$^{+1.1}_{-0.7}$ & 1.4$^{+0.2}_{-0.4}$  & 0.6$^{+0.3}_{-0.1}$\\
Ar$^{\ddag}$ = Ca$^{\ddag}$    & 
1.5$^{+0.7}_{-0.6}$  & 2.0$^{+1.1}_{-0.8}$ & 4.6$^{+1.0}_{-2.0}$  & 1.1$^{+0.3}_{-0.3}$\\
Fe$^{\ddag}$ = Ni$^{\ddag}$    & 
3.3$^{+3.5}_{-2.0}$  & 4.5$^{+6.1}_{-3.1}$  & $<$0.01 & $<$0.2 \\
Others$^{\ddag}$  & 1 (fixed)  & 1 (fixed)  & 1 (fixed) & 1 (fixed)\\
$\chi^2$/d.o.f. &
354/193 &  249/191 & 248/192 & 226/192 \\
\hline \\ [-6pt]
\multicolumn{5}{c}{Spectra-b}\\
\hline \\ [-6pt]
$N_{\rm H}$ ($\times10^{22}$ cm$^{-2}$)  &
1.7$^{+0.3}_{-0.2}$   & 2.1$^{+0.3}_{-0.3}$ & 1.5$^{+0.1}_{-0.1}$ &2.1$^{+0.3}_{-0.1}$ \\
$kT_{\rm e}$ (keV)  & 
0.93$^{+0.05}_{-0.03}$  & 0.27$^{+0.03}_{-0.05}$ & 0.39$^{+0.04}_{-0.02}$  & --- \\
   & ---     & 1.00$^{+0.06}_{-0.04}$ &     ---  & ---  \\
$kT_{\rm z}$/$kT_{\rm e}$ & ---  & --- & 2.8$^{+0.3}_{-0.2}$ & ---\\
$kT_{\rm e1}$ (keV)  &  --- & --- & --- & 5 (fixed)\\
$kT_{\rm e2}$ (keV)  &  --- & --- & --- & 0.30$^{+0.05}_{-0.04}$\\
$n_{\rm e}t^{\dag}$ ($\times10^{11}$ cm$^{-3}$ s) & 
---  & --- & --- & 4.8$^{+1.2}_{-0.1}$\\
Mg$^{\ddag}$   & 
0.6$^{+0.2}_{-0.1}$  & 0.7$^{+0.1}_{-0.2}$ & 0.4$^{+0.2}_{-0.2}$ & 0.5$^{+0.2}_{-0.2}$\\
Si$^{\ddag}$   & 
0.4$^{+0.1}_{-0.1}$  & 0.5$^{+0.2}_{-0.1}$ & 0.8$^{+0.2}_{-0.1}$  & 0.6$^{+0.1}_{-0.1}$\\
S$^{\ddag}$     & 
0.8$^{+0.2}_{-0.1}$ & 1.0$^{+0.2}_{-0.2}$ & 1.5$^{+0.4}_{-0.3}$ & 0.6$^{+0.1}_{-0.2}$\\
Ar$^{\ddag}$ = Ca$^{\ddag}$    & 
1.7$^{+0.5}_{-0.4}$  & 1.6$^{+0.5}_{-0.4}$ & 4.0$^{+1.0}_{-1.5}$  & 1.0$^{+0.5}_{-0.3}$\\
Fe$^{\ddag}$ = Ni$^{\ddag}$    & 
0.2$^{+0.3}_{-0.2}$  & 0.2$^{+0.3}_{-0.2}$  & $<$0.01 & 0.3$^{+0.3}_{-0.2}$\\
Others$^{\ddag}$  & 1 (fixed)   & 1 (fixed) & 1 (fixed)& 1 (fixed) \\
$\chi^2$/d.o.f. &
423/193 & 373/191 & 306/192 & 268/192 \\
\hline\\
\end{tabular}
\end{center}
\vspace{-10pt}
$^{\ast}$ Errors are estimated at the 90\% confidence levels.\\
$^{\dag}$ Recombination timescale, 
where $n_{\rm e}$ is the electron density (cm$^{-3}$) and $t$ is the elapsed time 
after the cooled down epoch (s).\\
$^{\ddag}$ Relative to the solar value \citep{Anders1989}.\\
\end{table*}

The spectra of G346.6$-$0.2 were made by subtracting BGD-a or BGD-b 
(here spectra-a and spectra-b, respectively). These spectra were separately given in figures 3a
and 3f, respectively.  
For these spectra, we applied a thin thermal plasma model 
in collisional ionization equilibrium (CIE) ({\tt vapec} model in XSPEC) 
modified by interstellar absorption ({\tt wabs} model in XSPEC).
The cross-sections of the photoelectric absorption were taken from \citet{Morrison1983}, 
while the abundance data were taken from \citet{Anders1989}.
The abundances of Mg, Si, S, Ar, and Fe were set to be free and 
those of Ca and Ni were assumed to be the same as Ar and Fe, respectively. 
The others were fixed to the solar values.
The XIS\,0, 1 and 3 spectra were simultaneously fitted.
The 1-CIE model was rejected with the large $\chi^2$ values of 354 and 423 (d.o.f.=193) 
for spectra-a and spectra-b, respectively.
The best-fit parameters are listed in table 1, while
the residuals from the best-fit model are plotted in figures 3b and 3g.
We found systematic residuals (shoulder-like structures) in the 1.5--4 keV energy band.  
We further examined a plasma model in an NEI state 
({\tt vnei} model in XSPEC) to the source spectra and confirmed that 
the model was also rejected with the $\chi^2$ values of 309 and 428 (d.o.f.=192) for 
spectra-a and spectra-b, respectively. 
The residuals were essentially similar to those of 1-CIE model.
We note that similar shoulder-like structures are found in the NEI model fit of 
\citet{Sezer2011} (figure 3), 
although the data reduction process was different from the present paper.

We then applied a 2-component CIE plasma (2-CIE) model (2-{\tt vapec} model), 
assuming the abundances of the two plasma components to be equal. 
This model was also rejected with the $\chi^2$ values of 249 and 373 (d.o.f.=191) 
for spectra-a and spectra-b, respectively (table 1).
The residuals from  the best-fit model are plotted in figures 3c and 3h. 
The residuals in the 1.5--2 keV  band disappeared, but 
those at the energy of 2--4 keV remained. 
These residuals show shoulder-like shapes with the leading edge energies of $\sim$2.4 keV 
and $\sim$3.2 keV which correspond the ionization energies of He-like Si and He-like S ions, 
respectively.
The same features were also reported from IC443, G359.1$-$0.5, W28, and W44  
\citep{Yamaguchi2009, Ohnishi2011, Sawada2012, Uchida2012}. 
These authors predicted that the residuals are due to 
the RRC which is a sign of the RP.

We, therefore, applied a {\tt cie} model in the SPEX package \citep{Kaastra1996} 
applying the {\tt absm} model for the interstellar absorption.
The {\tt cie} model treats an electron temperature ($kT_{\rm e}$) and an ionization temperature 
($kT_{\rm z}$) independently (here 1-RP model).
The best-fit parameters are listed in table 1.
Although the $\chi^2$ value for spectra-b was significantly improved from 373 
(d.o.f=191) to 306 (d.o.f=192), that for spectra-a was not improved (248 of d.o.f=192).  
Furthermore, the best-fit abundances of Ar and Ca are 4--5 solar, 
which is far higher than any other elements of solar or sub-solar abundances.  
The residuals from the best-fit 1-RP model are plotted in figures 3d and 3i.  
The residuals at 2--4 keV are partly reduced, but a new residual was appeared at the energy 
of $\sim$1.35 keV corresponding to the He-like Mg K$\alpha$ line.

The additional excess from the He-like Mg K$\alpha$ and unreasonable high abundances of Ar and Ca 
suggest that the single ionization-temperature (1-$kT_{\rm z}$) plasma in the {\tt cie} model is inadequate.   
In fact, \citet{Sawada2012} and \citet{Uchida2012} found that the spectra of MM SNRs, 
W28 and W44, have multi-ionization temperatures. 
They successfully predicted the multi-ionization temperature structure with a scenario 
that the X-ray emitting plasma is in a transition phase of recombining process.
In order to examine a possibility of a multi-$kT_{\rm z}$ plasma for G346.6$-$0.2, 
we fitted the spectra with the {\tt neij} model in SPEX.
The {\tt neij} model describes a plasma state when the initial plasma was in CIE with the temperature of
$kT_{\rm e1}$, then only the electron temperature dropped to $kT_{\rm e2}$ by a rapid electron cooling 
in the past. 
In the initial phase, the plasma was in 1-$kT_{\rm z}$ plasma ($kT_{\rm z}$=$kT_{e1}$), 
then forced recombination as a function of $n_{\rm e} t$, 
where $n_{\rm e}$ and $t$ are an electron density and an elapsed time after the electron cool-down epoch, 
respectively. 
Since the time sale of recombination is element-dependent, the time evolution realizes a multi-$kT_{\rm z}$ 
plasma. 

The {\tt neij} model gave better reduced $\chi^2$ value than the simple 
1-RP ({\tt cie}) model with $\Delta \chi^2$ = 22 (spectra-a) and 38 (spectra-b), 
but no constraint on $kT_{\rm e1}$ was obtained with the error ranges of
$>$4 keV and $>$3 keV  (90\% confidence level) for spectra-a and spectra-b, respectively. 
These error ranges mean that the initial plasma of $kT_{\rm e1}$ is in nearly full ionized state 
for all the relevant elements except for Fe (Fe would be in between  full ionized  and H-like states).  
Taking account of the initial plasma condition, we assumed $kT_{\rm e1}$=5~keV as a physically 
reasonable value and tried the final {\tt neij} fitting. 
The best-fit parameters and the $\chi^2$ values are listed in table~1. 
The best-fit model is plotted in figures~3a and 3f, while the residuals are shown in figures~3e and 3j. 
On the contrary to the simple 1-RP model, 
the {\tt neij} model gave reasonable abundances of Ar and Ca similar to the other elements.
Thus, we conclude that the overall spectra of G346.6$-$0.2 were nicely fitted with the {\tt neij} model.

\subsubsection{Local Emission around G346.6$-$0.2}

In the background estimation process, we found a local excess in soft X-rays around G346.6$-$0.2.
The X-ray spectrum of this local excess emission was made by subtracting BGD-b from the BGD-a data 
(figure 2), which is given in figure 4.
The X-ray spectrum shows a weak Si K$\alpha$ line, which means a thin thermal origin.
The spectrum was nicely fitted by a thin thermal plasma model, the {\tt apec} in XSPEC,
with $\chi^2$/d.o.f. =102/104.
The results are listed in table 2 and the best-fit model is plotted in figure 4.

\begin{figure}
  \begin{center}
   \FigureFile(8cm,8cm){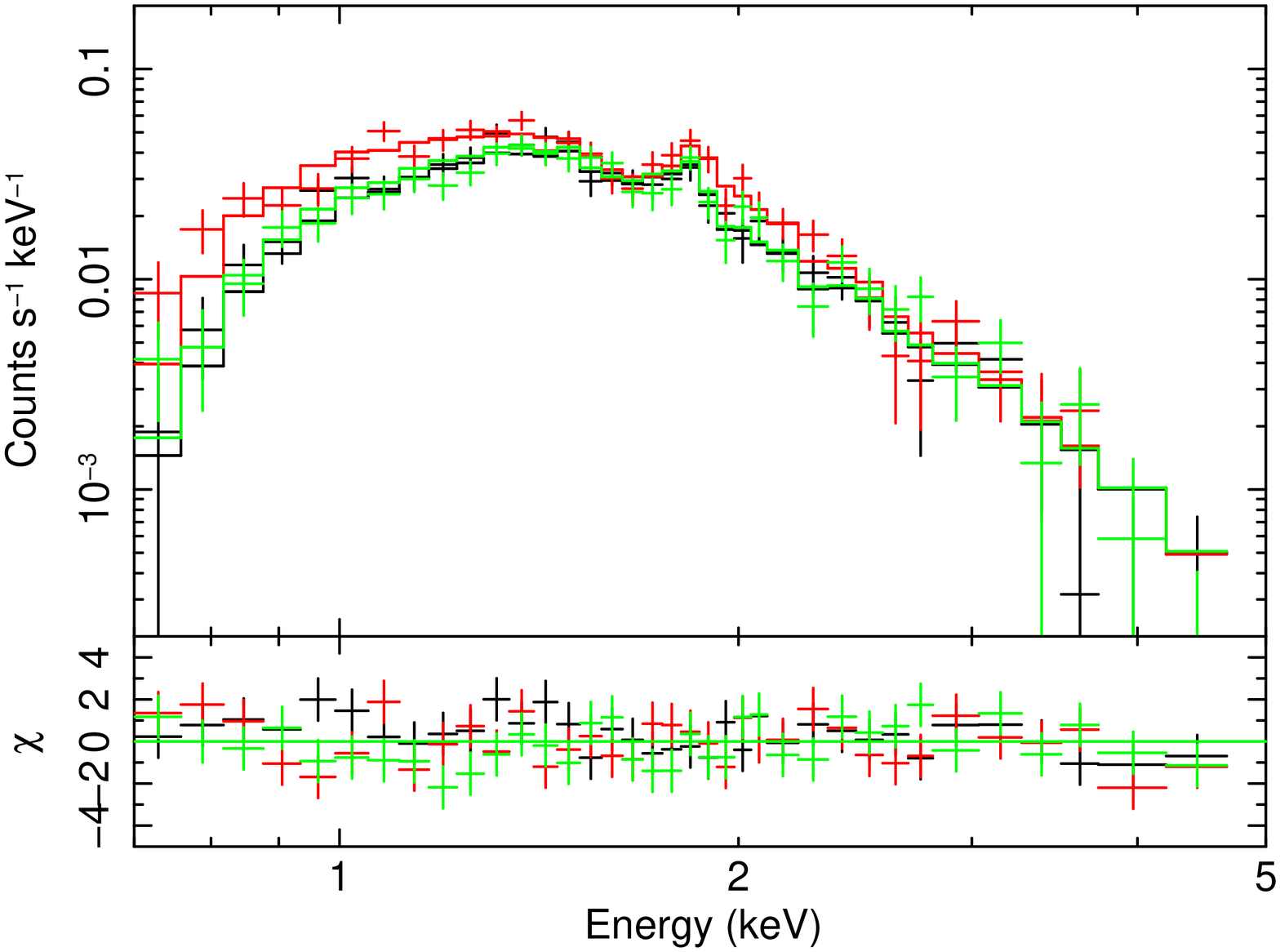}
  \end{center}
  \caption{
An X-ray spectrum of an excess emission around G346.6$-$0.2. 
The black, red, and green colors show XIS\,0, 1, and 3 data, respectively.
The best-fit {\tt apec} model is plotted by the solid histogram.
 }\label{fig:sample}
\end{figure}

\begin{table}[t]
\begin{center}
\caption{The best-fit parameters derived from a spectral analysis for an excess emission 
around G346.6$-$0.2.$^{\ast}$}
\end{center}
\begin{center}
\begin{tabular}{lc} \hline  \\ [-6pt]
Parameter & Value \\
Model &  XSPEC: apec\\
\hline \\ [-6pt]
$N_{\rm H}$ ($\times10^{22}$ cm$^{-2}$)  & 1.2$^{+0.1}_{-0.2}$\\
$kT_{\rm e}$ (keV)  & 0.79$^{+0.06}_{-0.05}$\\
Abundance$^{\dag}$    & 0.14$^{+0.06}_{-0.04}$ \\
$\chi^2$/d.o.f. & 102/104\\
\hline\\
\end{tabular}
\end{center}
\vspace{-10pt}
$^{\ast}$ Errors are estimated at the 90\% confidence levels.\\
$^{\dag}$ Relative to the solar value \citep{Anders1989}.\\
\end{table}

\section{Discussion}

\subsection{G346.6$-$0.2}

For the background estimation, we took account of the difference of vignetting effects 
between the source and background regions. 
Furthermore, we tried two different backgrounds of BGD-a and BGD-b 
(spectra-a and spectra-b). 
 Although the fluxes of these backgrounds are different, the spectral shapes are very similar. 
As the results, these two spectra, spectra-a and spectra-b, gave essentially the same best-fit 
parameters except for the absolute luminosity.
Since BGD-a and G346.6$-$0.2 are likely in a local soft X-ray excess around  G346.6$-$0.2, 
we discuss based on the results of spectra-a, the BGD-a subtracted spectra.

The spectrum was nicely described by the RP 
in a transition epoch of its recombining phase ({\tt neij} model). 
The best-fit electron temperature was about 0.3~keV. 
These conclusions are inconsistent with those of \citet{Yamauchi2008} and \citet{Sezer2011}. 
In the thermal plasma fitting, their results were in CIE or NEI with the electron temperatures of 
$\sim$1.6 keV and 0.97--1.3 keV in \citet{Yamauchi2008} and \citet{Sezer2011}, respectively. 
These apparent inconsistencies are mainly due to the background subtraction. 
\citet{Sezer2011} used the background from very small region (near the corner of the FOV), 
while \citet{Yamauchi2008} used the annulus region around the source.  
No vignetting effect was corrected in \citet{Yamauchi2008} and \citet{Sezer2011}.
Therefore, the background-subtracted spectra provide less statistics and 
should contain some fractions of the GRXE emission, particularly in the hard X-ray band. 
In fact, they reported that the X-ray spectrum has excess flux above $\sim$5 keV.
This may artificially predict higher electron temperatures and/or power-law component.

Assuming the mean density of 1H cm$^{-3}$, the derived $N_{\rm H}$ value of 
(2.3$\pm0.1$)$\times$10$^{22}$ cm$^{-2}$ corresponds to the distance of 7--8 kpc,
which is well consistent with the distance of 8.2 kpc estimated from $\Sigma$-$D$ relation \citep{Case1998}. 
If we assume the distance of 8 kpc, the luminosity was calculated to be 3.5$\times10^{35}$ erg s$^{-1}$ 
in the 0.5--10 keV energy band.
The abundances are sub-solar to solar, which is consistent with those of \citet{Sezer2011}. 
They predicted that the remnant may originate from a Type Ia supernova (SN) explosion, 
and  the solar/sub-solar abundances of heavy elements are explained by the scenario that the SNR is 
young, and the reverse shock does not reach yet to the interior of the ejecta.
Our best-fit $n_{\rm e} t$ is (4.8$^{+0.1}_{-0.4}$)$\times10^{11}$ cm$^{-3}$ s. 
Then, assuming the mean density of 1H cm$^{-3}$, the age of G346.6$-$0.2 is 
estimated to be 1.4$\times10^4$--1.6$\times10^4$ yr.
This means that G346.6$-$0.2 is not a young SNR, and hence no evidence for the young Type Ia SN 
scenario is obtained. The solar/sub-solar abundances plasma would be mainly due to 
interstellar matter.

G346.6$-$0.2 is the sixth sample of the SNRs with the RP, after IC443 \citep{Yamaguchi2009}, W49B 
\citep{Ozawa2009}, G359.1$-$0.5 \citep{Ohnishi2011}, W28 \citep{Sawada2012}, and W44
\citep{Uchida2012}. Among these SNRs with the RP,
the plasma structures in W28 and W44 were studied in detail. 
The results were that the RP is in a recombining phase of $\sim 10^{11}$($n_{\rm e}/1{\rm H~cm}^{-3}$)$^{-1}$~s 
after the production of the initial RP.  We found that G346.6$-$0.2 is explained with 
a similar scenario as these SNRs. 
We propose that the other SNRs, IC443, W49B, and G359.1$-$0.5, can also be explained by the same 
scenario, a plasma in an epoch of recombining process. 

All the previous RP-detected SNRs share some common characteristics: 
(1) they are classified to MM SNRs \citep{Rho1998}, 
(2) an OH maser is detected, suggesting the interaction with molecular clouds, and
(3) TeV/GeV $\gamma$-ray emission is detected. 
For possible origins of the RP based on these common features, one can refer the discussions in 
\citet{Yamaguchi2009}, \citet{Sawada2012}, and \citet{Uchida2012}.
The center-filled thermal X-ray emission within the radio 
shell suggests that G346.6$-$0.2 is a MM SNR.
An OH maser has been found \citep{Green1997}, but 
no TeV/GeV $\gamma$-ray emission has been found from G346.6$-$0.2.  
Therefore, future TeV/GeV $\gamma$-ray search from this SNR is encouraged. 

\subsection{Local Emission }

The excess emission around G346.6$-$0.2 was shown by the optically thin thermal plasma with the 
temperature of 0.79 keV, 0.14 solar abundance, and 
$N_{\rm H}$ value of 1.2$\times10^{22}$ cm$^{-2}$.
The ASCA Galactic plane survey revealed that the GRXE spectra observed 
in various regions are well represented by an optically thin thermal plasma model with
two temperatures of $<$1 keV and $\sim$7 keV \citep{Kaneda1997a,Kaneda1997b}. 
The two-temperature structure was confirmed by Chandra \citep{Ebisawa2005} and Suzaku 
\citep{Ryu2009}.
The spectral parameters of the excess emission in the G346.6$-$0.2 field
are similar to those of the soft component of the GRXE.
The intensity distribution of the hard component along the Galactic plane is 
symmetric with respect to the Galactic center, while that of the soft component is 
more asymmetric: a local peak and a local minimum were 
found at $l\sim$347$^{\circ}$ and $l\sim$355$^{\circ}$, respectively \citep{Kaneda1997a}; 
the longitudinal distribution of the GRXE soft component shows that the intensity near the 
BGD-a region is about $\sim$1.3 times higher than that of the BGD-b region \citep{Kaneda1997a}. 
Our present result of BGD-a is $\sim$2 times larger than BGD-b  in the soft X-ray band, 
and hence the excess would not be due to the fluctuation of the GRXE, but would be a local plasma. 
The $N_{\rm H}$ value of the local plasma is smaller than that of G346.6$-$0.2 in the GRXE, 
hence this emission is located at the near side of G346.6$-$0.2 and the GRXE.
The local plasma is near to the direction of the non-thermal SNR RX J1713.7$-$3946 and 
the $N_{\rm H}$ value is similar to that of RX J1713.7$-$3946 \citep{Koyama1997}.
Therefore, the local plasma is located near RX J1713.7$-$3946 at the distance of 1--2 kpc. 
Whether this thermal plasma is physically associated with RX J1713.7$-$3946 or not is unclear. 
To clarify this, we encourage further observations around RX J1713.7$-$3946.

\bigskip

We would like to express our thanks to all of the Suzaku team. 
We thank Dr. M. Sawada for his useful comments.
This work was supported by
the Japan Society for the Promotion of Science (JSPS); the Grant-in-Aid 
for Scientific Research (C) 21540234 (SY), 24540232 (SY), and 24540229 (KK), 
Young Scientists (B) 24740123 (MN),
Challenging Exploratory Research program  20654019 (KK), 
and Specially Promoted Research 23000004 (KK).



\begin{thebibliography}{}
\bibitem[Anders \& Grevesse(1989)]{Anders1989}
   Anders, E., \& Grevesse, N. 1989, Geochim. Cosmochim. Acta, 53, 197

\bibitem[Case \& Bhattacharya(1998)]{Case1998}
  Case, G. L., \& Bhattacharya, D. 1998, \apj, 504, 761
  
\bibitem[Clark et al.(1975)]{Clark1975}
   Clark, D. H., Caswell, J. L., \& Green, A. J. 1975, 
   Australian Journal of Physics Astrophysical Supplement, Sept., 1

\bibitem[Dubner et al.(1993)]{Dubner1993}
   Dubner, G. M., Moffett, D. A., Goss, W. M., \& Winkler, P. F.
   1993, \aj, 105, 2251
   
\bibitem[Ebisawa et al.(2005)]{Ebisawa2005}
   Ebisawa, K., et al. 2005, \apj, 635, 214
   
\bibitem[Green(2009)]{Green2009}
   Green, D. A. 2009, A Catalog of Galactic Supernova Remnants 
   (2009 March version), 
   (Astrophysics Group, Cavendish Laboratory, Cambridge, UK) 
   \footnote{http://www.mrao.cam.ac.uk/surveys/snrs/}

\bibitem[Green et al.(1997)]{Green1997}
   Green, A. J., Frail, D. A., Goss, W. M., \& Otrupcek, R. 1997, 
   \aj, 114, 2058

\bibitem[Hyodo et al.(2008)]{Hyodo2008}
  Hyodo, Y., Tsujimoto, M., Hamaguchi, K., Koyama, K., Kitamoto, S., Maeda, Y., Tsuboi, Y., \& Ezoe, Y.
  2008, \pasj, 60, S85
  
\bibitem[Kaastra et al.(1996)]{Kaastra1996}
  Kaastra, J. S., Mewe, R., \& Nieuwenhuijzen, H. 1996, UV and X-ray Spectroscopy of Astrophysical 
  and Laboratory Plasmas, ed. K. Yamashita and T. Watanabe (Universal Academy Press, Tokyo), p.411

\bibitem[Kaneda (1997)]{Kaneda1997a}
   Kaneda, H. 2007, Ph.D. thesis, The University of Tokyo
  
\bibitem[Kaneda et al.(1997)]{Kaneda1997b}
   Kaneda, H., Makishima, K., Yamauchi, S., Koyama, K., Matsuzaki, K., \& Yamasaki, N. Y. 
   2007, \apj, 491, 638
   
\bibitem[Koyama et al.(1986)]{Koyama1986}
   Koyama, K., Makishima, K., Tanaka, Y., \& Tsunemi, H. 1986, \pasj, 38, 121

\bibitem[Koyama et al.(1997)]{Koyama1997}
   Koyama, K., Kinugasa, K., Matsuzaki, K., Nishiuchi, M., Sugizaki, M., Torii, K., Yamauchi, S., \& 
   Aschenbach, B. 1997, \pasj, 49, L7
   
\bibitem[Koyama et al.(2007)]{Koyama2007}
   Koyama, K., et al.\ 2007, \pasj, 59, S23

\bibitem[Mitsuda et al.(2007)]{Mitsuda2007}
   Mitsuda, K., et al.\ 2007, \pasj, 59, S1

\bibitem[Morrison \& McCammon(1983)]{Morrison1983}
   Morrison, R., \& McCammon, D. 1983, ApJ, 270, 119

\bibitem[Nakajima et al.(2008)]{Nakajima2008}
   Nakajima, H., et al. 2008, \pasj, 60, S1

\bibitem[Ohnishi et al.(2011)]{Ohnishi2011}
  Ohnishi, T., Koyama, K., Tsuru, T. G., Masai, K., Yamaguchi, H., \& Ozawa, M. 
  2011, \pasj, 63, 527
  
\bibitem[Ozawa et al.(2009)]{Ozawa2009}
   Ozawa, M., Koyama, K., Yamaguchi, H., Masai, K., \& Tamagawa, T.
   2009, \apj, 706, L71
   
\bibitem[Rho \& Petre(1998)]{Rho1998}
  Rho, J., \& Petre, R. 1998, \apj, 503, L167
  
\bibitem[Ryu et al.(2009)]{Ryu2009}
   Ryu, S., Koyama, K., Nobukawa, M., Fukuoka, R., \& Tsuru, T. G 2009, \pasj, 61, 751
   
\bibitem[Sawada \& Koyama(2012)]{Sawada2012}
  Sawada, M., \& Koyama, K., 2012, \pasj, in press
  
\bibitem[Serlemitsos et al.(2007)]{Serlemitsos2007}
   Serlemitsos, P., et al.\ 2007, \pasj, 59, S9

\bibitem[Sezer et al.(2011)]{Sezer2011}
   Sezer, A., G\"{o}k, F., Hudaverdi, M., Kimura, M., \& Ecran, E. N. 2011, \mnras, 415, 301

\bibitem[Tawa et al.(2008)]{Tawa2008}
   Tawa, N., et al. 2008, \pasj, 60, S11
   
\bibitem[Uchida et al.(2012)]{Uchida2012}
   Uchida, H., et al. 2012, submitted to PASJ
   
\bibitem[Uchiyama et al.(2009)]{Uchiyama2009}
   Uchiyama, H., et al. 2009, \pasj, 61, S9
   
\bibitem[Whiteoak \& Green(1996)]{Whiteoak1996}
   Whiteoak, J. B. Z., \& Green A. J. 1996, \aaps, 118, 329 

\bibitem[Yamaguchi et al.(2009)]{Yamaguchi2009}
   Yamaguchi, H., Ozawa, M., Koyama, K., Masai, K., Hiraga, J., Ozaki, M., \& Yonetoku, D.
   2009, \apj, 705, L6
   
\bibitem[Yamauchi \& Koyama(1993)]{Yamauchi1993}
   Yamauchi, S. \& Koyama, K. 1993, \apj, 404, 620
   
\bibitem[Yamauchi et al.(2008)]{Yamauchi2008}
   Yamauchi, S., Ueno, M., Koyama, K., \& Bamba, A.
   2008, \pasj, 57, 459
\end{thebibliography}
\end{document}